\documentclass[12pt]{article}

\usepackage{scicite}

\usepackage{times}

\usepackage{caption}

\usepackage{graphicx}

\usepackage{amsmath}

\usepackage{xcolor} 

\newenvironment{sciabstract}{%
\begin{quote} \bf}
{\end{quote}}

\title{Attosecond timing of electron emission from a molecular shape resonance}

\author{
\parbox{\linewidth}{\centering
S. Nandi$^{1,2,\ast,\dagger}$, E. Pl\'{e}siat$^{3,\dagger}$, S. Zhong$^{1,\dagger}$, A. Palacios$^{3,4}$, D. Busto$^{1}$, M. Isinger$^{1}$, L. Neori\v{c}i\'{c}$^{1}$, C. L. Arnold$^{1}$, R. J. Squibb$^{5}$, R. Feifel$^{5}$, P. Decleva$^{6}$, A. L'Huillier$^{1}$, F. Mart\'{i}n$^{3,7,8,9}$, M. Gisselbrecht$^{1}$}
\\
\\
\normalsize{$^{1}$Department of Physics, Lund University, 22100 Lund, Sweden}\\
\normalsize{$^{2}$Universit\'{e} de Lyon, Universit\'{e} Claude Bernard Lyon 1, CNRS,}\\
\normalsize{Institut Lumi\`{e}re Mati\`{e}re, F-69622, Villeurbanne, France}\\
\normalsize{$^{3}$Departamento de Qu\'{i}mica, M\'{o}dulo 13,}\\ \normalsize{Universidad Aut\'{o}noma de Madrid, 28049 Madrid, Spain}\\
\normalsize{$^{4}$ Institute of Advanced Research in Chemical Sciences (IAdChem),}\\ \normalsize{Universidad Aut\'{o}noma de Madrid, 28049 Madrid, Spain}\\
\normalsize{$^{5}$Department of Physics, University of Gothenburg, 41296 G\"{o}teborg, Sweden}\\
\normalsize{$^{6}$Dipartimento di Scienze Chimiche e Farmaceutiche,}\\
\normalsize{Universit\'{a} di Trieste and IOM-CNR, 34127 Trieste, Italy}\\
\normalsize{$^{7}$Instituto Madrile\~{n}o de Estudios Avanzados en Nanociencia}\\
\normalsize{(IMDEA-Nanociencia), Cantoblanco, 28049 Madrid, Spain}\\
\normalsize{$^{8}$Condensed Matter Physics Center (IFIMAC),}\\
\normalsize{Universidad Aut\'onoma de Madrid, 28049 Madrid, Spain}\\
\normalsize{$^{9}$Donostia International Physics Center (DIPC),}\\
\normalsize{Paseo Manuel de Lardizabal 4, 20018 Donostia-San Sebasti\'an, Spain}\\
\\
\normalsize{$^\dagger$These authors contributed equally to this work.}
\\
\normalsize{$^\ast$Corresponding author, E-mail: saikat.nandi@univ-lyon1.fr}
}

\date{}

\begin{document}

\baselineskip24pt

\maketitle 

\begin{sciabstract}
Shape resonances in physics and chemistry arise from the spatial confinement of a particle by a potential barrier. In molecular photoionization, these barriers prevent the electron from escaping instantaneously, so that nuclei may move and modify the potential, thereby affecting the ionization process. By using an attosecond two-color interferometric approach in combination with high spectral resolution, we have captured the changes induced by the nuclear motion on the centrifugal barrier that sustains the well-known shape resonance in valence-ionized N$_2$. We show that despite the nuclear motion altering the bond length by only $2\%$, which leads to tiny changes in the potential barrier, the corresponding change in the ionization time can be as large as $200$ attoseconds. This result poses limits to the concept of instantaneous electronic transitions in molecules, which is at the basis of the Franck-Condon principle of molecular spectroscopy.

\end{sciabstract}

\section*{Introduction}

Shape resonances, due to trapping of particles in potential barriers, are ubiquitous in nature. First discovered by Fermi when studying slow-neutron capture in artificial radioactivity \cite{fermi1934,bohr1936}, they have since been the focus of countless investigations in physics, chemistry and biology. They play a crucial role in alpha-decay of radioactive nuclei \cite{blatt1979}, molecular fragmentation \cite{stolte2001}, rotational predissociation \cite{lovejoy1990}, electron detachment \cite{andersen1996}, ultracold collisions \cite{henson2012}, low-energy electron scattering \cite{schulz1973,palmer1992}, and photoionization \cite{dehmer1979,piancastelli1999}, to name a few. They are also thought to be at the origin of enhanced radiation damage of DNA and other biomolecules \cite{martin2004} and to play an important role in the stability of Bose-Einstein condensates \cite{yao2019}. Shape resonances are usually associated to specific spectral features. For instance, they lead to broad peaks in the photoionization spectrum of atoms and molecules \cite{west1980,dehmer1984,haack2000,plesiat2019}, and to a strong variation in the corresponding photoelectron angular distribution as a function of kinetic energy \cite{adachi2003}.

The energies at which these resonances are expected to appear and the corresponding trapping times (or conversely, decay lifetimes) are entirely determined by the shape of the barrier seen by the trapped (or ejected) particle, hence the name ``shape resonances''. 
Thus, the analysis of the resonance peaks observed in experimental spectra can be used to infer the actual height and width of the potential barrier seen by the impinging or emitted particle. This can be unambiguously done in atomic systems. However, in molecules the situation is more complicated, as nuclear motion may cause the potential felt by the electrons to change during a vibrational period, thus affecting the shape of the barrier. This was theoretically predicted back in 1979 by Dehmer, who argued that the anomalous variations of vibrationally-resolved photoelectron angular distributions of N$_2$ with photon energy could be the consequence of changes in the shape of the potential barrier with the internuclear distance \cite{dehmer1979}. This prediction has been the subject of debate for years \cite{piancastelli1999}. So far, direct time-resolved measurements of the dynamical evolution of shape resonances as the molecule vibrates have not been possible due to the lack of temporal resolution. 

In this work, we investigate with attosecond time resolution the changes induced by the vibrational motion on the potential barrier that sustains the $3\sigma_g^{-1}$ shape resonance in N$_2$. We do so by ionizing the molecule with extreme ultraviolet (XUV) radiation consisting of high-order harmonics spanning an energy range between $20$ and $40$ eV [see Fig.\ref{fig:fig1}(a)]. Temporal information is obtained by using the RABBIT (Reconstruction of Attosecond Beating By Interference of Two-photon transitions) interferometric method, described below \cite{paul2001}. In particular, we determine the (relative) photoionization delay, which is the time the electron takes to escape from the molecular potential, and is thus sensitive to the presence of a shape resonance. With the support of theoretical calculations that explicitly describe the photoionization process and take into account the vibrational motion, we show that changes in the molecular bond length of only $0.02$ \AA~ can lead to variations of the photoionization delay as large as $200$ attoseconds, and are not uniform over the investigated energy range. These measurements with attosecond time resolution show that molecular photoionization in the vicinity of shape resonances cannot be described in terms of the commonly accepted Franck-Condon picture, in which electronic excitation is assumed to be instantaneous and decoupled from the nuclear motion.

\section*{Results and Discussion}
\begin{figure*}[!htb]
    \centering
    \includegraphics[width=1.0\textwidth]{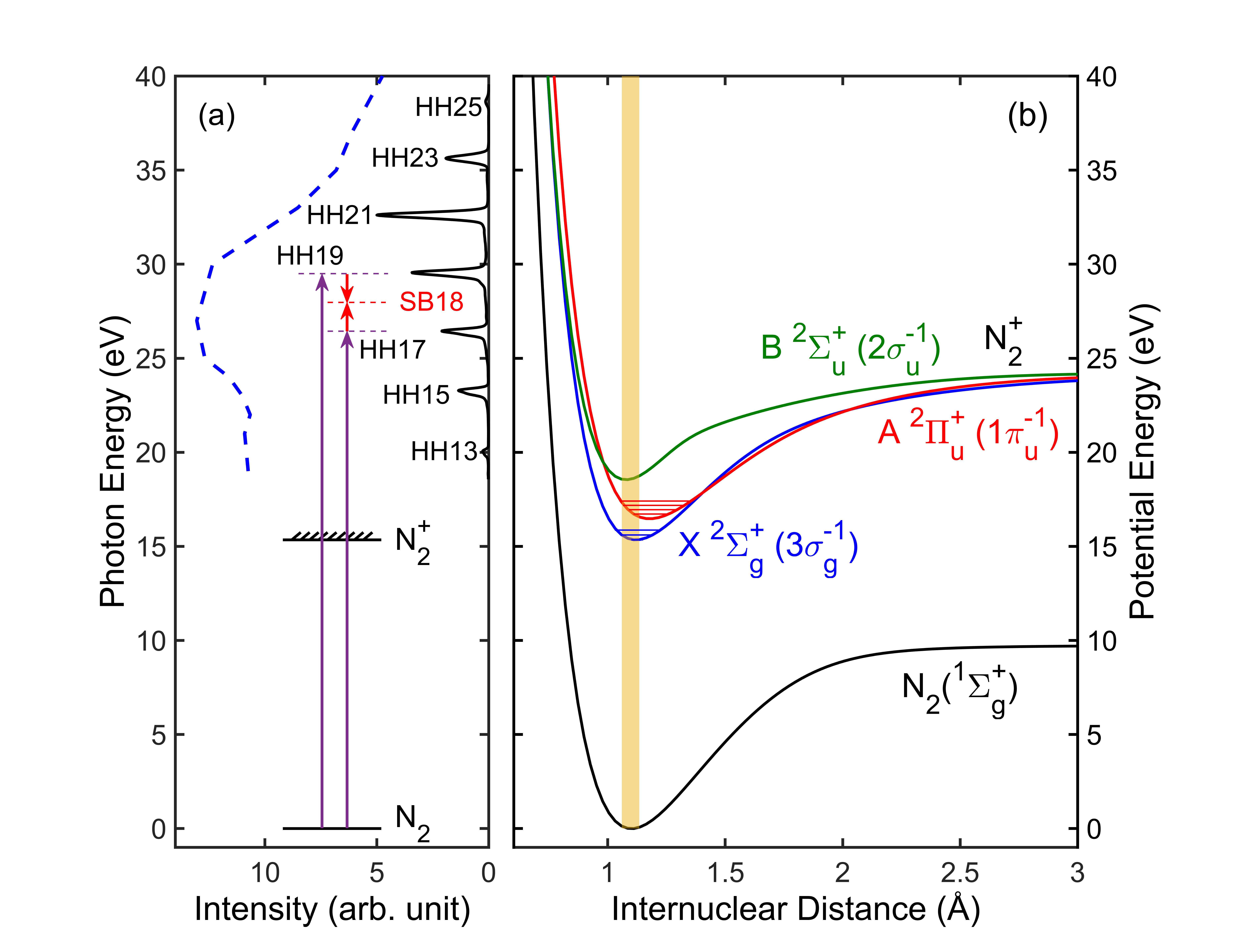}
    \caption{\label{fig:fig1} \textbf{Photoionization scheme.} (a) A comb of high-order harmonics (HH), spanning over the photon energy range of $20-40$ eV, probes the entire shape resonance region in N$_2^+$. The blue dashed line corresponds to experimental photoionization cross-sections [taken from \cite{hamnett1976}] across the resonance. The scheme for generating sideband (SB) $18$ is denoted by the vertical arrows. The violet (red) arrows denote XUV (NIR) photons. 
    (b) Potential energy curves of the ground state of N$_2$: $^1\Sigma_g^+$ (black), and lowest three states of N$_2^+$: $X~^2\Sigma_g^+$ (blue), $A~^2\Pi_u^+$ (red) and $B~^2\Sigma_u^+$ (green). The shaded area ($1.06-1.14$ \AA) denotes the Franck-Condon region. The horizontal lines show the position of different vibrational levels associated with the corresponding electronic state.}
    \end{figure*}

The RABBIT technique has been widely utilized to determine accurate photoionization delays in atoms \cite{klunder2011,isinger2017} and, following the pioneering work of Haessler \textit{et al.} \cite{haessler2009}, has started to be applied to molecular systems as well \cite{huppert2016,cattaneo2018,vos2018}. 
Here, N$_2$ is ionized by a comb of odd high-order harmonics, covering the $3\sigma_g^{-1}$ shape resonance 
[see Fig.\ref{fig:fig1}(a)]. Electrons are mainly removed from the $3\sigma_g$ and $1\pi_u$ orbitals of the N$_2$ molecule in its $^1\Sigma_g^+$ ground state, leading to N$_2^+$ ions in the $X~^2\Sigma_g^+$ and $A~^2\Pi_u^+$ states ($X$ and $A$ states for short), respectively [see Fig.\ref{fig:fig1}(b)].

In our experiment (see Materials and Methods for details), a near infrared (NIR) $45$ fs laser pulse generated high-order harmonics in argon, corresponding to a train of attosecond pulses in the time domain \cite{antoine1996}. The harmonics and a weak replica of the NIR pulse (probe) were focused into an effusive gas jet containing N$_2$ molecules. The ejected electrons were detected by a magnetic bottle electron spectrometer with resolution up to $E/\Delta E \sim 80$. To utilise the best possible resolution and hence resolve the vibrational levels in the first two outer valence states of the N$_2^+$ ion, photoelectrons were retarded by suitable voltages before entering into the spectrometer flight tube. 
The measurements consisted in recording, for alternating shots, the XUV-only and XUV+NIR photoelectron spectra (PES) as a function of the delay ($\tau$) between the XUV and NIR pulses. 

We perform simulations to explicitly obtain the vibrationally resolved photoelectron spectra resulting from the interaction of an isolated N$_2$ molecule with an XUV attosecond pulse train and a time-delayed NIR pulse, so that the extracted time delays can be directly compared with experiment. The time-dependent Schr\"{o}dinger equation is numerically solved including the bound-bound, bound-continuum and continuum-continuum dipole transition matrix elements between the electronic states. These are computed using the static exchange density functional theory method described in \cite{plesiat2018}. The nuclear motion is taken into account within the Born-Oppenheimer approximation (see Theoretical Methods in Supplementary Materials for a detailed explanation) and the laser parameters are chosen to reproduce the experimental conditions.

The measured and calculated PES, obtained with XUV and NIR, exhibit sidebands originating from the interference between two quantum paths: the absorption of a harmonic and an NIR photon, and the absorption of the next harmonic and the stimulated emission of an NIR photon [see Fig.\ref{fig:fig1}(a)]. 
Consequently, the amplitude of the sidebands, $A_\mathrm{SB}$, oscillates as a function of $\tau$ according to the formula \cite{paul2001}: 
\begin{equation}\label{rabbit}
    A_\mathrm{SB}=\mathcal{A}+\mathcal{B}~\mathrm{cos}[2\omega_0( \tau-\tau_\mathrm{XUV}-\tau_\mathrm{mol})],
\end{equation}
where $\mathcal{A}$ and $\mathcal{B}$ are two constants, $\omega_0$ the angular frequency of the driving NIR field, $\tau_\mathrm{XUV}$ denotes the group delay of the attosecond pulses \cite{mairesse2003} and $\tau_\mathrm{mol}$ is the molecular two-photon ionization time delay  \cite{baykusheva2017} (see Supplementary Materials). The combination of both high spectral and temporal resolution achieved in our experiment allows us to distinguish photoelectrons leaving the residual molecular cation in different vibrational states, and to determine the variation in photoionization delays due to changes in the molecular geometry.

\begin{figure*}[!htb]
    \centering
    \includegraphics[width=1.0\textwidth]{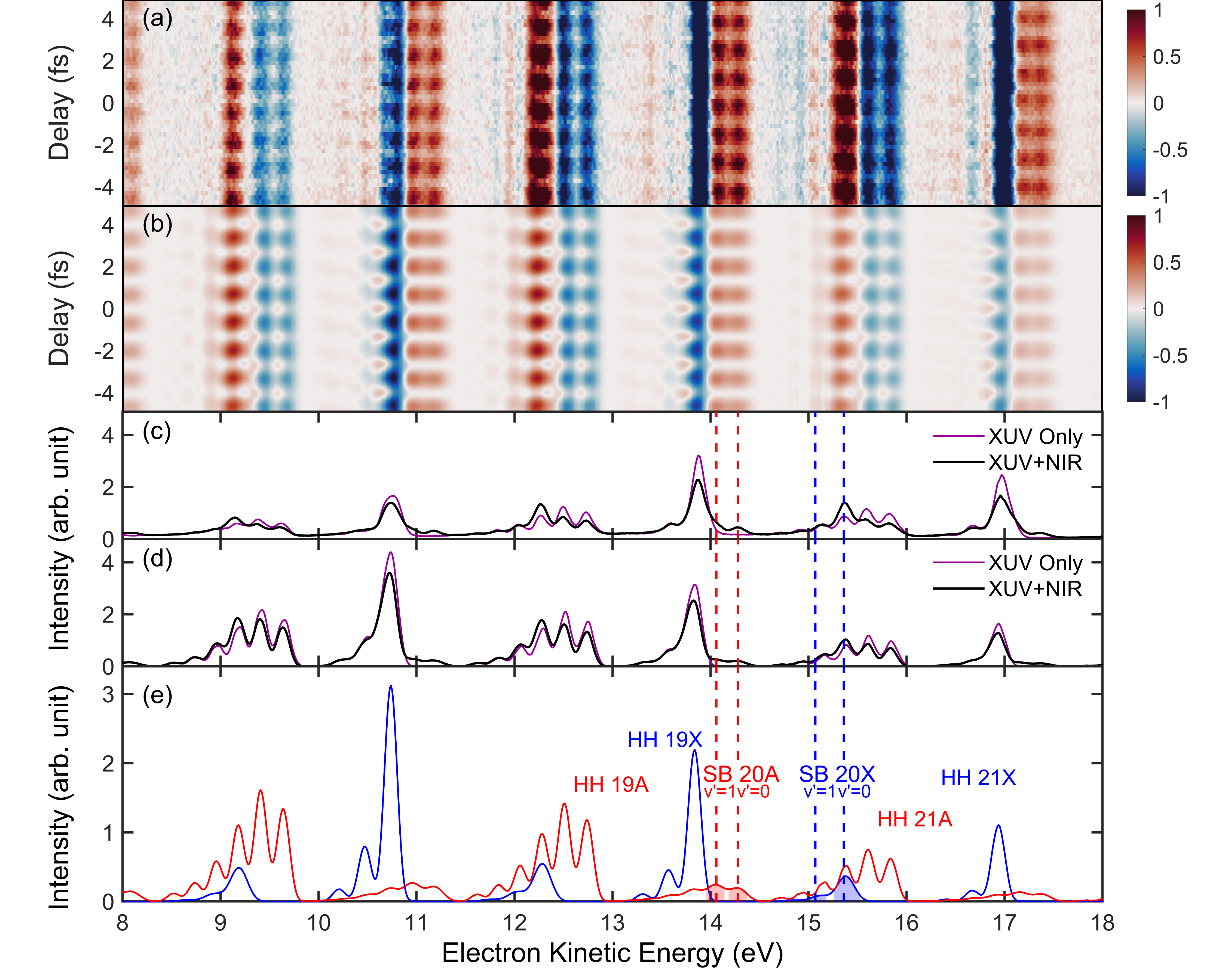}
    \caption{\label{fig:fig2} \textbf{Photoelectron Spectra.} Difference between photoelectron spectra obtained with XUV+NIR and XUV-only, as a function of delay. Experiment (a) and Theory (b). Experimental (c) and theoretical (d) photoelectron spectra for XUV only (violet) and XUV+NIR (black) photoionization, averaged over all relative delays. 
    (e) Theoretical photoelectron spectra for XUV+NIR photoionization to the $X$ (blue) and $A$ (red) electronic states. By comparing (a) and (e), we can assign the spectral features to different vibrational levels (shaded area) of the $X$ and $A$ electronic states, as indicated by the vertical blue and red dashed lines.}
    \end{figure*}

\begin{figure*}[!htb]
    \centering
    \includegraphics[width=1.0\textwidth]{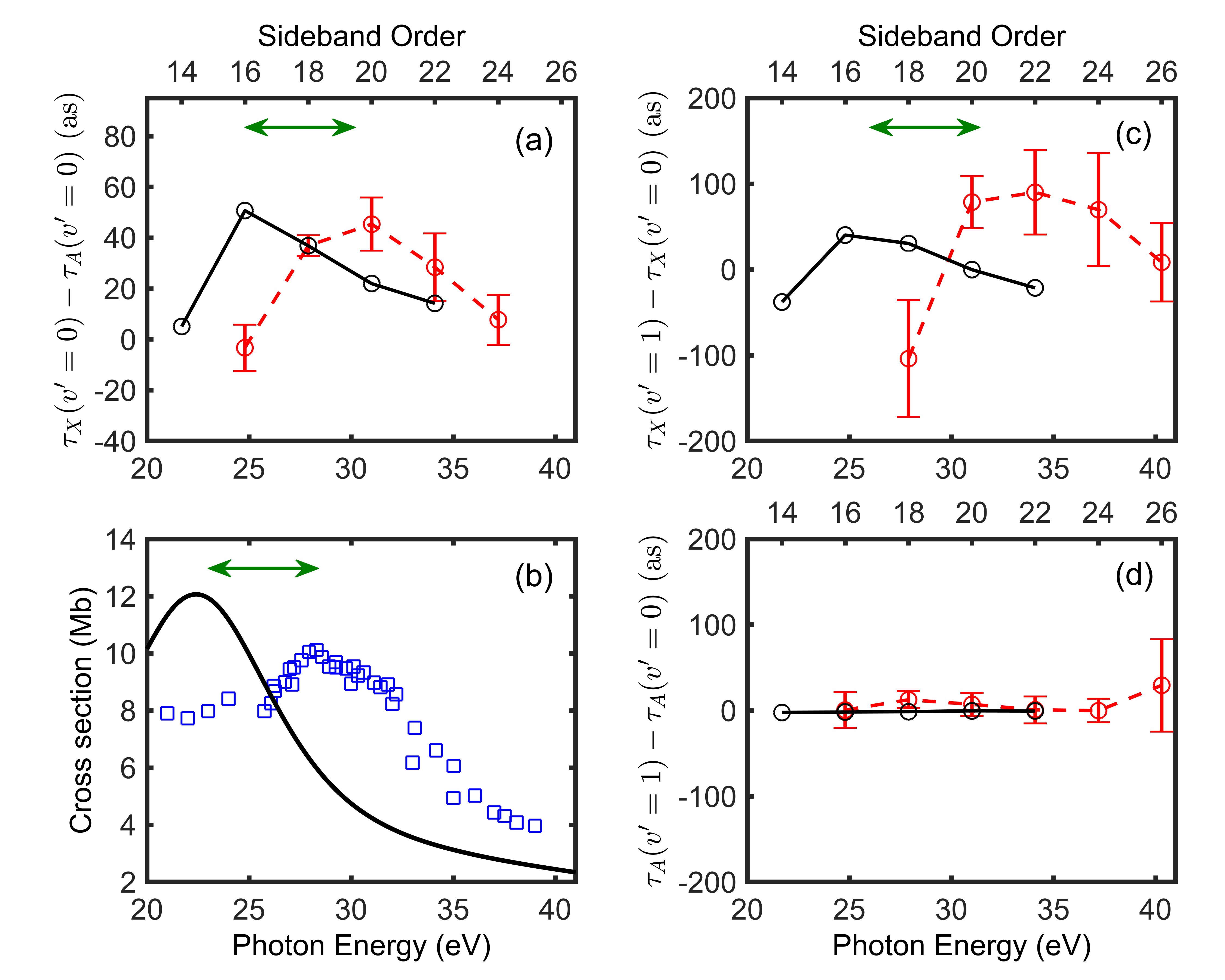}
    \caption{\label{fig:fig3} \textbf{Photoionization time-delays.} (a) Differences in molecular time delays $\left(\tau_{mol}\right)$ between $X$- and $A$-states for $v'=0$. Red circles: experiment, black circles: theory. (b) Partial photoionization cross-sections for the $X$-state; open-square: synchrotron-based experiment \cite{hamnett1976,plummer1977}, solid line: theory (this work). The position of the resonance-maxima is shifted by almost $6$ eV (denoted as $\leftrightarrow$) between theory and experiment. 
    This shift is also observed in the relative time delay [(a),(c)]. (c) Relative time delay between the vibrational levels $v'=1$ and $0$ for the $X$-state. The strong photon energy dependence observed here vanishes completely if one neglects the nuclear motion (see Fig.S4 in SM). (d) Same as (c), but for the $A$-state.  }
    \end{figure*}

Fig.\ref{fig:fig2}(a) and Fig.\ref{fig:fig2}(b) show the experimental and theoretical data, respectively, corresponding to the difference between the XUV+NIR and XUV-only PES (in color), which oscillates with frequency $2\omega_0$ as a function of $\tau$. Both theory and experiment are in excellent agreement (for details about the theoretical method, see Supplementary Materials). Fig.\ref{fig:fig2}(c) presents the XUV-only (violet) and XUV+NIR (black) PES obtained by integrating over all delays. Good agreement with the calculated spectra shown in Fig.\ref{fig:fig2}(d) can be noticed. The difference between theory and experiment in the relative intensities of some of the photoelectron peaks is due to the different position of the shape resonance predicted by theory [see Fig.\ref{fig:fig3}(b)]. In addition, the harmonic comb used in the theoretical calculations was slightly different from the experimental one. Fig.\ref{fig:fig2}(e) presents individual contributions from the $X$ (blue) and $A$ (red) states in the theoretical XUV+NIR PES, which allows us to assign the different features of the experimental spectra. The structures between $14.5$ and $16$ eV, for example, are due to ionization to the $A$-state by absorption of the 21\textsuperscript{st} harmonic leaving the N$_2^+$ ion in the $v'=0-6$ vibrational states and to a two-photon transition (sideband) to the $X$-state with vibrational states $v'=0,1$ (blue shaded region). The small peaks near $14$ eV are due to two-photon transitions to the $A$-state with vibrational states $v'=0,1$ (red shaded region). Plotting the difference between the XUV+NIR and XUV-only PES allows us to confirm our assignment, since, as shown in Fig.\ref{fig:fig2}(a), we can distinguish between sideband (positive) and harmonic peaks (negative). 

To determine molecular two-photon ionization time delays $\tau_\mathrm{mol}$, we fitted the measured [Fig.\ref{fig:fig2}(a)] sideband oscillations to Eq.~(\ref{rabbit}). The same procedure is applied to extract the theoretical ones from the RABBIT spectra computed with a time-dependent numerical approach. Fig.\ref{fig:fig3}(a,c,d) show experimental (red circles) and theoretical (black circles) relative time delays for different final states. Since the contribution from the ionizing radiation ($\tau_\mathrm{XUV}$) is the same for all the final states of N$_2^+$, the plotted differences correspond to pure molecular contributions.   

The relative molecular time delay for leaving N$_2^+$ in the $X(v'=0)$ state with respect to leaving it in the $A(v'=0)$ state is shown in Fig.\ref{fig:fig3}(a). For both theory and experiment, this relative delay varies by more than $40$ attoseconds across the shape resonance region. The theoretical results, however, are shifted to lower energies with respect to the experimental ones by almost $6$ eV, indicated by the green arrow in Fig.\ref{fig:fig3}(a), which is similar to the shift of the maximum of the calculated photoionization cross-section in comparison with that obtained from synchrotron radiation measurements \cite{hamnett1976,plummer1977} [see Fig.\ref{fig:fig3}(b)]. This is due to an incorrect description of the resonance position by our theoretical method, which is necessarily simpler than state-of-the-art electronic structure methods for the equilibrium geometry \cite{plesiat2012}, as we must describe the molecular electronic continuum states in a wide range of internuclear distances, as well as all the continuum-continuum transitions induced by the NIR probe pulse (see Supplementary Materials).

Fig.\ref{fig:fig3}(c) and (d) show the relative molecular time delays, $\tau_X\left(v'=1\right)-\tau_X\left(v'=0\right)$ and $\tau_A\left(v'=1\right)-\tau_A\left(v'=0\right)$, for the $X$ and $A$ electronic states respectively. For the $A$-state, the relative delay is very small and practically independent of photon energy, while for the $X$-state, it varies significantly across the shape resonance. Once again, for the reasons described above, the theoretical curve is shifted down in energy with respect to the experimental one by $\sim 6$ eV.  

The variation of the molecular time delay differences between the $X$- and $A$-states observed in Fig.\ref{fig:fig3}(a) is therefore mainly due to the variation of the time delay for the former, which can be attributed to the presence of the shape resonance. The time delay varies with energy because the time spent by the photoelectron in the metastable state before being ejected into the continuum also varies with energy. Since for the $A$-state the electron does not have to go through any potential barrier, the corresponding time delay is much smaller than for the $X$-state.

We now analyze the physical meaning of the results presented in Fig.\ref{fig:fig3}(c,d). In atomic systems, photoionization time delays obtained from RABBIT measurements can often be written as the sum of two contributions, $\tau_1+\tau_{cc}$. The first term is related to one-photon ionization by the XUV field. For a single or dominant ionization channel containing no sharp structures in the continuum (e.g., narrow Fano-resonances), $\tau_1$ is given by the derivative of the scattering phase in that particular channel, the so-called Wigner delay \cite{wigner1955,smith1960}. The second term, $\tau_{cc}$, is the additional time delay due to the continuum-continuum transitions induced by the NIR field  \cite{dahlstrom2011,dahlstrom2012}.

\begin{figure*}[!htb]
    \centering
    \includegraphics[width=1.0\textwidth]{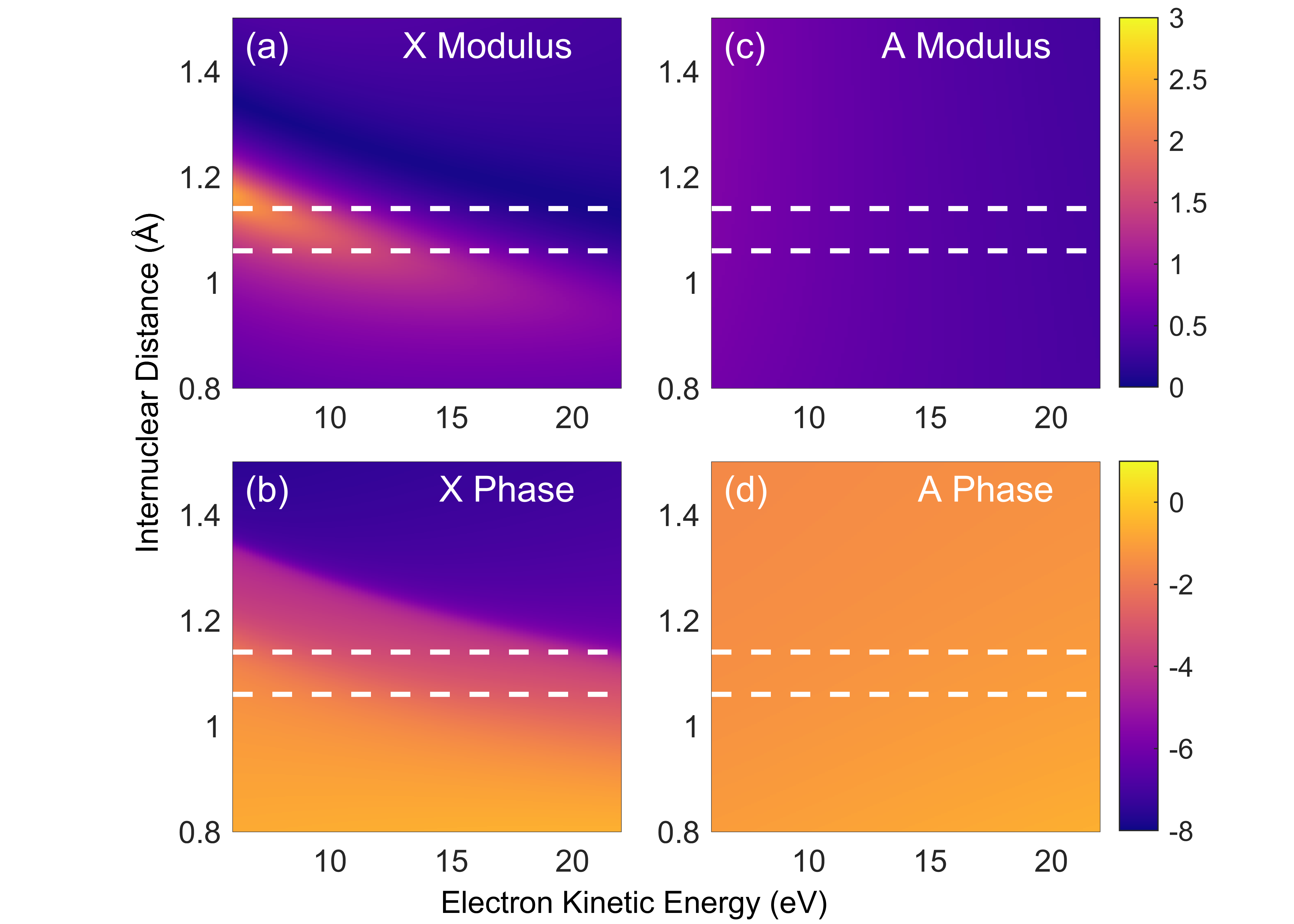}
    \caption{\label{fig:fig4} \textbf{One-photon dipole transition matrix elements.} Modulus (a,c) and phase (b,d) of the dominant terms $d^X_{\sigma_u,l=3,z}$ (a,b) and $d^A_{\delta_g,l=2,x}$ (c,d) contributing to the one-photon transition matrix element of the X- and A-states, respectively, as a function of the internuclear distance $\left(R\right)$ and electron kinetic energy $\left(\varepsilon\right)$. The area between the two dashed lines denotes the Franck-Condon region. }
    \end{figure*}

In the vicinity of the shape resonance, one-photon ionization leading to N$_2^+$ in the $X$-state is dominated by the $f$-wave ($\ell=3$). Similarly, for the $A$-state, the $d$-wave ($\ell=2$) dominates over all other partial waves in the same photon energy region (see Fig.S2 in Supplementary Materials). Although the molecular two-photon ionization time delay $\tau_\mathrm{mol}$ cannot be strictly decomposed as  $\tau_1+\tau_{cc}$, due to averaging over molecular orientation and electron emission angle, the variation of  $\tau_\mathrm{mol}$ in Fig.~\ref{fig:fig3}(c,d) still reflects the ionization dynamics arising from the main channels \cite{baykusheva2017}. Fig.~\ref{fig:fig4} shows the modulus and phase of the dominant terms contributing to the dipole transition element as a function of electron kinetic energy $\varepsilon$ and internuclear distance $R$ for both electronic states (see Supplementary Materials for notations). For the $X$-state, at a given $\varepsilon$, both the modulus [Fig.\ref{fig:fig4}(a)] and phase [Fig.\ref{fig:fig4}(b)] of the dipole transition element strongly vary with $R$ within the Franck-Condon region, in contrast to the $A$-state [Fig.\ref{fig:fig4}(c) and (d)]. This implies that electronic transitions cannot be considered instantaneous relative to nuclear motion as assumed by the widely used Franck-Condon picture. Consequently, the molecular photoionization delays, obtained by taking the derivative of the phase of the dipole transition element, strongly depend on $R$. The difference in molecular time delays between the $v'=1$ and $0$ vibrational levels thus provides direct information on non Franck-Condon ionization dynamics, i.e., on how the nuclear motion affects the photoionization process. 
 
\begin{figure*}[!htb]
    \centering
    \includegraphics[width=1.0\textwidth]{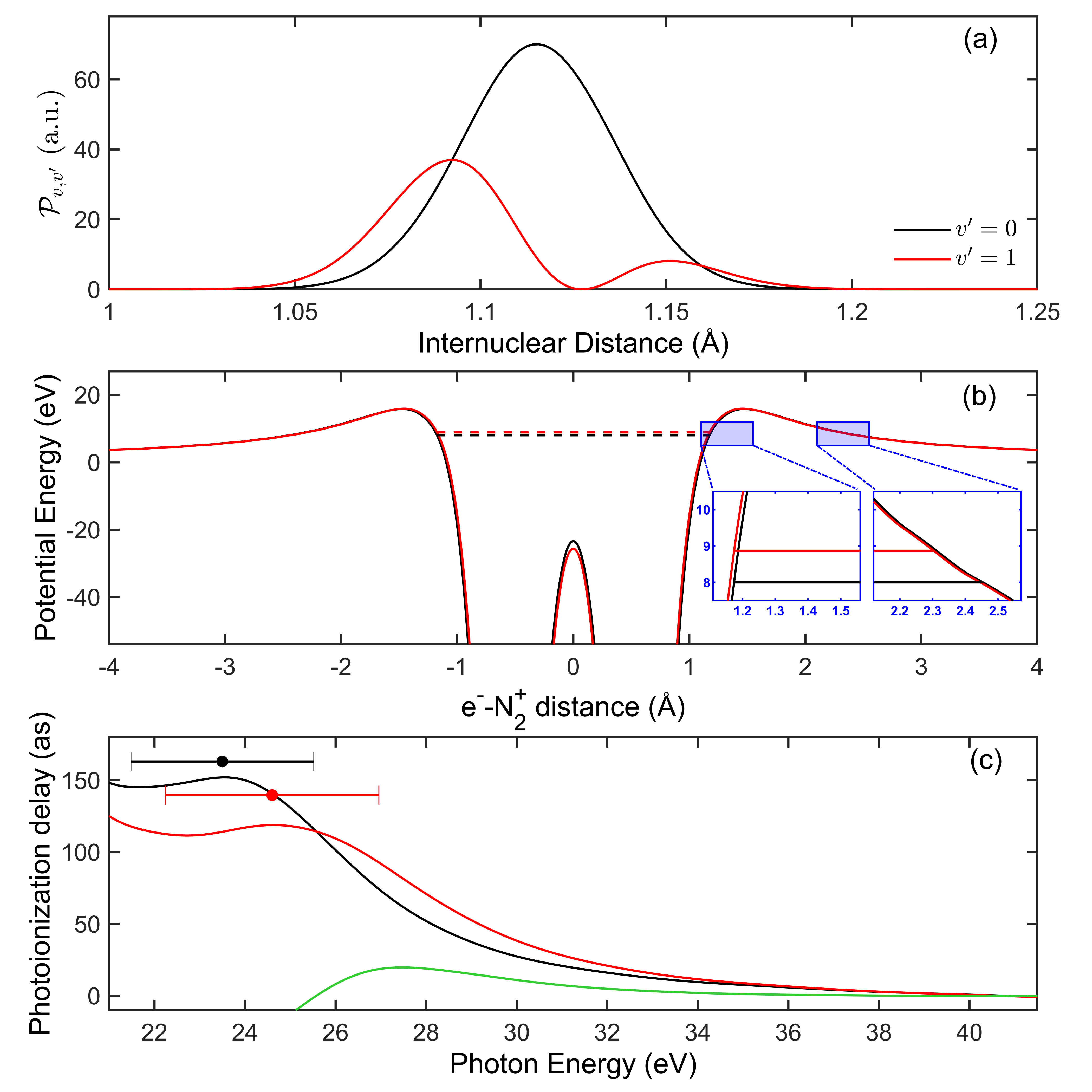}
    \caption{\label{fig:fig5} \textbf{Electron dynamics induced by structural changes} (a) Absolute square of the
transition matrix element $d^X_{\sigma_u,l=3,z}$ at an electron kinetic energy of $8.2$ eV, multiplied by the initial ($\chi_{i;v=0}$) and final ($\chi_{f;v'}$) vibrational wave-functions, $\mathcal{P}_{v,v'}$, as a function of the internuclear distance $R$. The red and the black curves correspond to transitions from the $v=0$ level in the neutral ground state to the $v'=1$ and the $v'=0$ levels of the $X$-state, respectively. (b) Potential felt by an electron escaping from an N$_2$ molecule with internuclear distance $R_{v'=1}=1.09$ \AA~ (red) and $R_{v'=0}=1.113$ \AA~ (black), calculated at the same of level of theory as the dipole matrix elements. These Kohn-Sham potentials are shown along the internuclear axis, with the two wells representing the two $N$-atoms. The origin is placed at the center of mass of the molecular ion. The dashed lines show the corresponding resonance energies.}
    \end{figure*}

Fig.\ref{fig:fig5}(a) shows the absolute square of the product between the initial vibrational wavefunction, the transition matrix element for the $X$-state [see Fig.\ref{fig:fig4}(a)] at an electron kinetic energy of $8.2$ eV, and the final vibrational wavefunction (see Supplementary Materials for details). The initial and final vibrational wavefunctions correspond, respectively, to the $v=0$ level of N$_2$ in the ground electronic state and the $v'=0$ (black) and $v'=1$ (red) levels of N$_2^+$ in the $X$ electronic state. These curves have well-defined maxima at $R=1.113$ \AA~ and $1.09$ \AA, respectively, showing that the transition to the shape resonance occurs, on average, at smaller internuclear distances for $v'=1$ than for $v'=0$. This small difference in bond length of $\sim 0.02$ \AA~ has a significant impact on the electron dynamics. Indeed, as illustrated in Fig.\ref{fig:fig5}(b), the potential felt by the emitted photoelectron at these two internuclear distances is different, leading to a higher resonance energy for $R=1.09$ \AA~ $\left(v'=1\right)$ than for $R=1.113$ \AA~ $\left(v'=0\right)$ as seen by comparing the red and black dashed lines. In addition, due to the different slopes on the rising [left inset in Fig.\ref{fig:fig5}(b)] and falling [right inset in Fig.\ref{fig:fig5}(b)] edges, the barrier is narrower and the resonance lifetime is shorter for $v'=1$ than for $v'=0$.

Fig.\ref{fig:fig5}(c) shows the photoionization delays resulting from the one-photon dipole transition matrix elements calculated at the two above-mentioned internuclear distances as a function of photon energy. The difference in internuclear distance leads to a noticeable shift in the position of the corresponding maxima of photoionization delays, in agreement with the difference in resonance energy discussed above as well as the positions of the maxima calculated using the Wentzel-Kramers-Brillouin (WKB) approximation, indicated by the black and red dots. In addition, the energy range where the photoionization delays vary significantly is slightly broader for $v'=1$ than for $v'=0$, a direct consequence of the shorter lifetime for $v'=1$. This can also be seen by the horizontal bars, representing the widths of resonance obtained with the WKB approximation. Both effects, due to the shape resonance, contribute to a variation of the time-delay difference between the $v'=1$ and $v'=0$ states as indicated by the green curve. Remarkably, such a simple model predicts the main features of the experimental results in Fig.\ref{fig:fig3}(c), in particular, the change of sign of the relative photoionization delay at low photon energy and the maximum at approximately the resonance position. Finally, it is worth noting that the resonance lifetimes obtained from the WKB model, $139.6$ and $163$ attoseconds for the $v'=1$ and $v'=0$ levels, respectively, are much smaller than the corresponding vibrational period, which is of the order of $16$ fs (see Supplementary Materials for details). As a consequence, the nuclei barely move during the ionization process, thus supporting the above analysis.

\section*{Conclusion}
 In summary, we measured vibrationally resolved molecular photoionization time delays between the $X$ and $A$ electronic states in N$_2$ across the $3\sigma_g^{-1}$ shape resonance using attosecond interferometry. This enabled us to capture the changes associated with nuclear motion on the centrifugal barrier seen by the escaping photoelectron. The observation of such changes goes beyond the usual Franck-Condon approximation which assumes that electronic transitions are instantaneous in comparison with nuclear motion. By combining attosecond time-resolution with high spectral resolution, we were able to break the temporal frontier beyond which the signature of the ``slow'' nuclear motion in molecular photoionization, can be seen and quantified. This approach should allow investigating the effect of nuclear motion on a variety of electronic processes in more complex molecular systems at the sub-femtosecond time scale.

\section*{Materials and Methods}
\paragraph*{Experimental Methods}

The output of a Ti:Sapphire laser system delivering NIR pulses around $800$ nm with $5$ mJ pulse energy at $1$ kHz repetition rate was sent to an actively stabilized Mach-Zehnder type interferometer. In the ``pump'' arm, the NIR pulses were focused into a gas cell containing argon atoms to produce a train of attosecond XUV pulses via high-order harmonic generation. A $200$ nm thick aluminum foil was used to filter out the co-propagating NIR pulse. The bandwidth of the driving NIR pulse was kept around $50$ nm which ensured the generation of high-order harmonics having full-width-at-half-maximum (FWHM) of about $150$ meV. This is significantly smaller than the energy separation ($267$ meV) between the lowest vibrational levels ($v'=1$ and $v'=0$) of the $X$ electronic state in N$_2^+$ (see Supplementary Materials), which is essential to vibrationally resolve the measured PES. In the ``probe'' arm, the NIR pulses could be delayed relative to the XUV pulses by a piezoelectric-controlled delay stage and blocked at each alternate shot by a mechanical chopper. After recombination, the collinearly propagating XUV and NIR pulses were focused by a toroidal mirror into an effusive gas jet of N$_2$. The emitted photoelectrons were detected by a  2~m-long magnetic bottle electron spectrometer, with a $4\pi$ solid angle collection. We estimated an average NIR intensity of $8\times10^{11}$ W/cm$^2$ in the interaction region.

\paragraph*{Data Analysis}
Due to the spectral congestion between ionization by XUV and XUV+NIR radiation to the three different states ($X$, $A$ and $B$) of the N$_2^+$ ion (see Fig.S1 in Supplementary Materials), a careful analysis of the experimental data was needed. Hence, a spectrally resolved variant of the RABBIT protocol, called Rainbow RABBIT \cite{gruson2016}, was used to analyze the experimental data. 
Despite the overlap between the $X$- and $B$-states, we could determine the phases of the sideband oscillations by carefully choosing the region with least possible spectral overlap. In addition, at relatively high photon energies ($>25$) eV, the photoionization cross-section to the $B$-state is much smaller than to the $X$-state (\textit{29}). Therefore the measured time-delays for the $X$-state can be considered to be effectively free from any possible spectral contamination from the $B$-state.

For every sideband, a Fourier transform was performed to make sure that the SB oscillation did not include frequency components higher than $2\omega_0$. 
The uncertainty $\sigma_X$ $\left(\sigma_A\right)$ for each measurement of the molecular photoionization time delay $\tau_X$ $\left(\tau_A\right)$ was obtained from the fit of the RABBIT oscillation to a cosine function [see Eq.(1), main text]. The corresponding uncertainty on the relative time delay, $\tau_{X}-\tau_{A}$, can be expressed as: 
\begin{equation}
    \sigma_{X-A}=\sqrt{\sigma_X^2+\sigma_A^2}.
\end{equation}
An identical procedure was used to calculate the uncertainties for the relative time delays between two vibrational levels of the same electronic state. 

The final experimental values shown in Fig.3 were obtained from a weighted average of the data points from several sets of measurements.  
For $N$ measurements yielding $N$ data points: $k_1,k_2,\ldots,k_N$ with corresponding uncertainties: $\sigma_1,\sigma_2,\ldots,\sigma_N$, the weighted average can be calculated as:
\begin{equation}
    \overline{k}=\frac{\sum_{i=1}^N w_i k_i}{\sum_{i=1}^N w_i},
\end{equation}
where $w_i=1/\sigma_i^2$ is the weight. The error bars indicated in Fig.3 are the weighted standard deviation, defined as
\begin{equation}
    \sigma_{\overline{k}}=\sqrt{\frac{N\sum_{i=1}^N w_i(k_i-\overline{k})^2}{(N-1)\sum_{i=1}^N w_i}}.
\end{equation}

\paragraph*{Funding} We acknowledge the support from the ERC advanced grant PALP-339253, the Swedish Research Council (grant no. 2013-8185), the Knut and Alice Wallenberg Foundation and the European COST Action AttoChem (CA18222). EP, AP and FM acknowledge the support of the MINECO project FIS2016-77889-R. FM acknowledges support from the `Severo Ochoa' Programme for Centres of Excellence in R\&D (MINECO, Grant SEV-2016-0686) and the `Mar\'{\i}a de Maeztu' Programme for Units of Excellence in R\&D (CEX2018-000805-M). AP acknowledges the support of a Ram\'on y Cajal contract (RYC-2014-16706) from Ministerio de Econom\'ia y Competitividad (Spain). EP acknowledges the support of a Juan de la Cierva contract (IJCI-2015-26997) from Ministerio de Econom\'ia y Competitividad (Spain). Calculations were performed at CCC-UAM and Marenostrum Supercomputer Center. 
\paragraph*{Acknowledgments} SN acknowledges fruitful discussions with V. Loriot and F. L\'{e}pine.
\paragraph{Author contributions} S.N. conceived the experiment, the planning for which was further improved by inputs from S.Z., A.L.H. and M.G.. S.N., S.Z., D.B., M.I., L.N., C.L.A. carried out the experiment. S.N. and S.Z. performed the data analysis. E.P. performed the theoretical calculations under the supervision of A.P., P.D. and F.M.. The model was developed by E.P., A.P. and F.M.. R.J.S. and R.F. provided the magnetic bottle electron spectrometer. M.G., A.L.H. and F.M. supervised the project. S.N., A.L.H., M.G. and F.M. wrote the manuscript with inputs from all the authors.
\paragraph{Competing interests} The authors declare no competing interests.
\paragraph{Data and materials availability} The data that support the findings of this study are included in the paper and/or, the Supplementary Materials. Any additional data related to this work are available from the corresponding author upon request. 

\clearpage

  \clearpage
    \begin{figure*}[!htb]
    \centering
    \caption*{As shown in the insets, the barrier width depends on $R$, hence, on the vibrational state of the ion.  (c) Photoionization delays $\left(\tau_1\right)$ as a function of photon energy obtained at the internuclear distances $R_{v'=1}=1.09$ \AA\ (red) and $R_{v'=0}=1.113$ \AA\ (black), and the relative delay between them, $\tau_1^X(R_{v'=1})-\tau_1^X(R_{v'=0})$ (green). The resonance lifetimes (full circles) and widths (horizontal bar) calculated in the WKB approximation from the two potentials presented in Fig.5(b) are also shown.}
    \end{figure*}

\section*{Supplementary Materials}

Theoretical Methods\\
Fig.S1. Theoretical XUV+NIR PES for \textit{X}, \textit{A} and \textit{B} states. \\
Fig.S2. Contributions from the different partial waves for \textit{X} and \textit{A} states. \\
Fig.S3. $\mathcal{P}_{v,v'}$ as a function of the internuclear distance $R$, for photoelectron kinetic energy of $11.7$ eV. Relative two-photon ionization delays between $v'=0$ (at $R=1.103$ \AA) and $v'=1$ (at $R=1.084$ \AA) level of the $X$-state as well as the one-photon ionization delays and their differences at the same internuclear distances.\\
Fig.S4. Comparison of theoretical time delays obtained with our method accounting for nuclear motion and within the Franck-Condon approximation. \\
References \textit{(38-42)}.

\end{document}


\baselineskip24pt

\maketitle

\section*{Theoretical Methods}
The theoretical method is based on the formalism described in (\textit{27}). We neglect molecular rotation, which is much slower than electronic and vibrational motions. For a given orientation of the molecular axis with respect to the polarization vector of the incident radiation, the time evolution of the wave function describing the interaction of the molecule with the light pulses is given (in atomic units) by the time-dependent Schr\"odinger equation (TDSE):
\begin{equation}
 \left[\hat{H}_0+\hat{V}\left(t\right)\right] \Psi\left(R,{\bf r},t\right) = i\frac{\partial{\Psi\left(R,{\bf r},t\right)}}{\partial{t}}
\end{equation}
where $\hat{H}_0$ is the field-free Hamiltonian of the molecule, $\Psi\left(R,{\bf r},t\right)$ the time-dependent wave function, which depends on the internuclear distance $R$, the coordinates ${\bf r}$ of the $N$ electrons and the time $t$. $\hat{V}\left(t\right)$ is the laser-molecule interaction potential, which reads, in the length gauge:
\begin{equation}
\hat{V}\left(t\right)= \mathbf{E}\left(t\right)\cdot\sum_{j=1}^{N}{\mathbf{r}_j},
\end{equation}
where $\mathbf{E}\left(t\right)$ is the electric field. In the region of the electronic continuum investigated in this work, one does not expect a significant contribution of non adiabatic effects. We thus assume that the time-dependent Born-Oppenheimer approximation is valid. This allows us to write the total wave function $\Psi\left(R,{\bf r},t\right)$ as a product of a nuclear vibrational wave function and a time-dependent electronic wave function that depends parametrically on the vibrational coordinate (\textit{38}), 
\begin{equation}
\Psi\left(R,{\bf r},t\right) = \Psi^{el}\left(R;{\bf r},t\right)\chi_{i,v}\left(R\right).
\end{equation}
As the response of the vibrational wave packet is much slower than that of the electrons and the considered time intervals are relatively short, we neglect the time dependence of the vibrational wave function, which implies that $\chi_{i,v}\left(R\right)$ is identical to the ground vibrational state of the molecule for all times.

Electron dynamics is entirely governed by the electronic time-dependent Schr\"odinger equation,
\begin{equation}
 \left[\hat{H}_0^{el}+\hat{V}\left(t\right)\right] \Psi^{el}\left(R;{\bf r},t\right) = i\frac{\partial{\Psi^{el}\left(R;{\bf r},t\right)}}{\partial{t}},
\end{equation}
where $\hat{H}_0^{el}$ is the field-free electronic Hamiltonian. The above equation depends parametrically on $R$ and, therefore, must be solved for all values of $R$ that are physically accessible.

We assume that the initial $N$-electron state of the molecule is well represented by a Slater determinant
\begin{equation}
\left|\Psi^{el}_i\left(R;{\bf r}\right)\right> = \frac{1}{\sqrt{N!}}{\rm det}\left\{ \left|\varphi_{i_1}\left(R;{\bf r}_{q_1}\right)\right>\left|\varphi_{i_2}\left(R;{\bf r}_{q_2}\right)\right>\cdots\left|\varphi_{i_N}\left(R;{\bf r}_{q_N}\right)\right>\right\} ,
\end{equation}
where $\varphi_{i_n}\left(R;{\bf r}_{q_j}\right)$ is the $n$-th spin-orbital for electron $j$.
As electron correlation is not expected to vary strongly during the short time intervals considered in our calculations, we further assume that the electronic time-dependent wave function is also described by a single Slater determinant:
\begin{equation}
\left|\Psi^{el}\left(R;{\bf r},t\right)\right> = \frac{1}{\sqrt{N!}}{\rm det}\left\{ \left|\psi_{1}\left(R;{\bf r}_{q_1},t\right)\right>\left|\psi_{2}\left(R;{\bf r}_{q_2},t\right)\right>\cdots\left|\psi_{N}\left(R;{\bf r}_{q_N,}t\right)\right>\right\},
\end{equation}
where $\psi_{k}\left(R;{\bf r}_{q_j},t\right)$ is the $k$-th time-dependent spin-orbital for electron $j$.
We are here assuming that the electronic Hamiltonian can be written as a sum of single-electron Hamiltonians:
\begin{equation}
\hat{H}_0^{el}=\sum_{j=1}^N\hat{h}^{(j)}_{KS}
\end{equation}
where $\hat{h}^{(j)}_{KS}=\frac{\hat{p}_j^2}{2}+\hat{v}_{KS}\left(R;{\bf r}_{q_j}\right)$ is the Kohn-Sham (KS) Hamiltonian describing the ground state of the molecule and $\hat{v}_{KS}$ is the Kohn-Sham potential that contains the usual Coulomb, exchange and correlation potentials. 
Within this approximation, the single-electron wave functions satisfy the single-particle time-dependent Schr\"odinger equation:  
\begin{equation}
 \left[\hat{h}^{(j)}_{KS}+\hat{v}^{(j)}\left(t\right)\right] \psi_{k}\left(R;{\bf r}_{q_j},t\right) = i\frac{\partial{\psi_{k}\left(R;{\bf r}_{q_j},t\right)}}{\partial{t}}
 \label{eq-tdse_2},
\end{equation}
where $\hat{v}^{(j)}\left(t\right)=\mathbf{E}\left(t\right)\cdot{\mathbf r}_j$ (in the length gauge).
We solved this equation by expanding the time-dependent $\psi_{k}$ spin-orbitals in a basis of field-free (stationary) KS spin-orbitals, 
\begin{equation}
 \psi_{k}\left(R;{\bf r}_{q_j},t\right) = \sum_{n}{c_{k,n}\left(t\right)\varphi_{n}\left(R;{\bf r}_{q_j}\right)e^{iE_n t}}+\sum_{q,l,\varepsilon}{c_{k,q,l,\varepsilon}\left(t\right)\varphi_{q,l,\varepsilon}\left(R;{\bf r}_{q_j}\right)e^{i\varepsilon t}}, \label{eq:spectral}
\end{equation}
and imposing the initial conditions $c_{k,n}=\delta_{k,i_k}$ and $c_{k,q,l,\varepsilon}=0$, for all initially occupied KS orbitals, i.e., $k=1, ..., N$. In the above expansion,
the functions $\varphi_{n}\left(R;{\bf r}_{q_j}\right)$ represent the bound stationary spin-orbitals with energy $E_n$, and $\varphi_{q,l,\varepsilon}\left(R;{\bf r}_{q_j}\right)$ the continuum stationary spin-orbitals of symmetry $q$, angular momentum $l$ and photoelectron energy $\varepsilon$. Bound and continuum spin-orbitals were obtained by solving the KS equations 
\begin{equation}
\hat{h}^{(j)}_{KS}\varphi_{k'}\left(R;{\bf r}_{q_j}\right)=E_{k'}\varphi_{k'}\left(R;{\bf r}_{q_j}\right).
\end{equation}
For the continuum spin-orbitals, we impose the usual incoming boundary conditions (\textit{39,27}). Substitution of Eq.~(\ref{eq:spectral}) in (\ref{eq-tdse_2}) leads to a system of coupled first-order differential equations that were solved numerically. 
 
As shown in previous works (\textit{40,27}), the $N$-electron probability describing the transition from the initial state $\Psi^{el}_i\left(R;{\bf r}\right)$ to a final state $\Psi^{el}_{q,l,\varepsilon}\left(R;{\bf r}\right)$ 
\begin{equation}
\left|\Psi^{el}_{q,l,\varepsilon}\left(R;{\bf r}\right)\right> = \frac{1}{\sqrt{N!}}{\rm det}\left\{ \left|\varphi_{f_1}\left(R;{\bf r}_{q_1}\right)\right>\left|\varphi_{f_2}\left(R;{\bf r}_{q_2}\right)\right>\cdots\left|\varphi^-_{q,l,\varepsilon}\left(R;{\bf r}_{q_N}\right)\right>\right\},
\label{eq:psi}
\end{equation}
written as a Slater determinant built from $N-1$ bound field-free KS orbitals and a continuum field-free KS orbital of energy energy $\varepsilon$, is given by 
\begin{equation}
{\Im}_{i\rightarrow f}^{q,l}\left(R,\varepsilon\right) = \left|\langle\Psi^{el}_{q,l,\varepsilon}\left(R;{\bf r}\right)|\Psi^{el}\left(R;{\bf r},t=T_{max}\right)\rangle\right|^2 ={\rm det}\left(\gamma^{q,l,\varepsilon}_{nn'}\right) \hspace{0.5cm} n, n' = 1, \cdots, N
\label{eq-pexc_1}
\end{equation}
where 
\begin{equation}
\gamma^{q,l,\varepsilon}_{nn'} =  \sum_{k=1}^N c^*_{k,f_n} (t=T_{max})\  c_{k,f_n'}(t=T_{max})
\end{equation}
and
$T_{max}$ is larger than or equal to the total pulse duration of the XUV and NIR radiation. In Eq.~(\ref{eq:psi}), The superscript $^-$ denotes the correct scattering boundary condition of the continuum state which has been included by means of the Galerkin approach (\textit{39}).

Finally, the vibrationally resolved $N$-electron transition probability is given by 
\begin{equation}
{\Im}_{i\rightarrow f}^{q,l,v,v'}\left(\varepsilon\right) = \left|\int{dR\ \chi_{f,v'}(R) \  \langle\Psi^{el}_{q,l,\varepsilon}\left(R;{\bf r}\right)|\Psi^{el}\left(R;{\bf r},t=T_{max}\right)\rangle\ \chi_{i,v}\left(R\right)}\right|^2
\label{eq-pexc_2}
\end{equation}
where $\chi_{i,v}\left(R\right)$ and $\chi_{f,v'}\left(R\right)$ are the initial and final vibrational wave functions. We emphasize that one has to calculate the above transition probability for all open ionization channels $\alpha$ leading to an electron with energy $\varepsilon$. For simplicity in the notation, we omit such index in all above expressions. 
In Eq.~(\ref{eq-pexc_2}) we have assumed that the nuclei do not move significantly during the absorption of the XUV and the NIR photons, therefore only the electronic part of the wave function is propagated in time. This is a reasonable approximation, as the NIR photon must be absorbed when the attosecond XUV pulse is still present.

The vibrational wave functions were obtained by solving the 1D vibrational Schr\"odinger equation by using the potential energy curves calculated with the CASSCF/MRCI method (\textit{32}). The calculated energy difference between $v'=1$ and $v'=0$ (equal to $267$ meV for the X channel and $230$ meV for the A channel), as well as the vibrational period of the ground state ($14.4$ fs), are in good agreement with the experimental values.
The KS orbitals were determined by diagonalizing the field-free KS Hamiltonian in a multicenter basis set of B-splines functions and symmetry-adapted real spherical harmonics (\textit{39}). The KS potential was built with the LB94 exchange-correlation functional (\textit{41}) and the electronic density was calculated by using the Amsterdam Density Functional package with a DZP basis set. The multicenter basis set consists of two types of expansion centers: one located at the center of mass of the molecule, with a large radius ($R^0_{max}=$1600 a.u.) and a large number of angular momenta ($l^0_{max}=$14), in order to describe accurately the long-range oscillatory behaviour of the continuum states, and two additional expansions centered on the nitrogen atoms  ($R^1_{max}=$0.6 a.u. and $l^1_{max}=$1) to improve the description of the cusp of the wave function in the vicinity of the nuclei.
The basis of KS orbitals was constructed for fifteen different values of the internuclear distance $R$ spanning from $1.55$ a.u. to $2.75$ a.u. (the equilibrium geometry of the ground state is $R_{eq}\sim$2.06 a.u.). Eq.~(\ref{eq-tdse_2}) was solved for each $R$ independently by using a Runge-Kutta propagator implemented in existing PETSC libraries. In our simulation, the duration of both the NIR pulse and the XUV train is $39$ fs. Their intensities are $5\times10^{11}$ W/ cm$^2$ and $5\times10^{10}$ W/cm$^2$, respectively. The central wavelength of the NIR pulse is $800$ nm. For the XUV, we have used a Gaussian HHG spectrum centered around $29.4$ eV with a FWHM$=14.4$ eV. In the temporal domain, it leads to a train of $29$ attosecond pulses separated by $1.33$ fs.

\paragraph*{Molecular photoionization time delays}
For linearly polarized light, we can define the time delay associated to one-photon ionization in the $\alpha$ channel as (\textit{29,33,34}):
\begin{align}
\tau^{\alpha}_{1}\left(R,\varepsilon\right) & =  
\frac{1}{8\pi^2}\int{d\Omega_M\int{d\Omega_e\frac{\sigma_{\alpha}\left(R,\varepsilon,\Omega_M,\Omega_e\right)}{\sigma_{\alpha}\left(R,\varepsilon\right)} }}
\label{eq-talpha}\\
& 
\times \frac{\partial}{\partial \varepsilon} \arg \left(\sum_{q,l,\mu}\left(-i\right)^{l}e^{i\eta_l\left(\varepsilon\right)}D^1_{\mu 0}\left(\Omega_M\right)d^\alpha_{q,l,\mu}\left(R,\varepsilon\right)\sum_m{b^q_{l, m}Y_{l}^{m}\left(\Omega_e\right)}\right)
\nonumber
\end{align}
where in the length gauge,
\begin{equation}
d^\alpha_{q,l,\mu}\left(R,\varepsilon\right) = \left<\varphi^-_{q,l,\varepsilon}\left(R;{\bf r}_{q_j}\right)\right|r_j Y_{1}^{\mu}\left|\varphi_{\alpha}\left(R;{\bf r}_{q_j}\right)\right>
\end{equation}
and $\Omega_M$ and $\Omega_e$ correspond respectively to the set of Euler angles of the polarization vector and emitted electron relative to the molecular frame. $\mu$ denotes the spherical component of the dipole operator, $\eta_l\left(\varepsilon\right)$ is the Coulomb phase shift, $D^1_{\mu 0}$ the Wigner rotation matrix for linearly polarized light, $b^q_{l, m}$ the coefficients of the angular expansion in real spherical harmonics $Y_{l}^{m}\left(\Omega_e\right)$. The quantities $\sigma_{\alpha}\left(R,\varepsilon,\Omega_M,\Omega_e\right)$ and $\sigma_{\alpha}\left(R,\varepsilon\right)$ are the angularly resolved and total photoionization cross sections respectively. 
Eq.~(\ref{eq-talpha}) is the average over electron emission angle and molecular orientation of the angle-dependent photoionization delay (\textit{42}). 

Additionally, we introduced the quantity, $\mathcal{P}_{v,v'}$, to examine the $R$-dependent photoionization yield for the $X$-state as: $\mathcal{P}_{v,v'} \left(R,\varepsilon\right)=\left|\chi_{i,v=0}\left(R\right)  d^X_{\sigma_u,l=3,z} \left(R,\varepsilon\right)
\chi_{f,v'}\left(R\right)\right|^2$.
This quantity allows us to identify the characteristic values of the internuclear distance at which ionization by the XUV photons leading to electrons with kinetic energy $\varepsilon$ is more likely. In Fig. 5, we have chosen an electron kinetic energy of $8.2$ eV, since this is the energy at which the shape resonance appears in our calculations. Meanwhile in Fig. S3 we have chosen an electron kinetic energy of $11.7$ eV, since in this case we are interested in understanding the behavior of the time delay difference observed in the experiment at higher energies.

\clearpage

Fig.S1 shows theoretical contributions from three electronic states: $X$ (blue), $A$ (red) and $B~^2\Sigma_u^+(2\sigma_u^{-1})$ (green), in the N$_2^+$ ion. 

Fig.S2 shows the contribution of the different partial waves for the $X$ (top) and $A$ (bottom) final states. Photoionization to the $X$-state is dominated by the contribution to the \textit{f}-wave ($\ell=3$, orange line) in the region of the shape resonance, while photoionization to the $A$-state is dominated by the transition to the \textit{d}-wave ($\ell=2$, light green line). 

Fig.S3(a) shows the same results as Fig.5(a), although obtained by analyzing $\mathcal{P}_{v,v'}$ at a higher  photoelectron kinetic energy, 11.7 eV. Fig.S3(b) shows the one-photon ionization delays and their corresponding differences for the $v'=1$ and $v'=0$ levels of the $X$-state, at the positions of the maxima in Fig.S3(a), $R=1.084$ \AA~ and $1.103$ \AA~, respectively. These values are very close to those of Fig.5(a), indicating that the photoelectron will experience essentially the same scattering potential on its way-out. Fig.S3(b) also presents the difference between the two-photon ionization delays for $v'=1$ and $v'=0$ at these internuclear distances. The good agreement obtained between the relative one-photon (green curve) and two-photon (blue circles) ionization delays is consistent with our interpretation of the experimental two-photon delays (see Fig.3, main text).

Fig.S4 shows a comparison between $\tau_{mol}$ presented in Fig.3(c) (black circles) and the relative two-photon time delays obtained by applying the Franck-Condon approximation (red squares). In the latter, the electronic part is assumed to be independent on the nuclear degrees of freedom, therefore the electronic coupling can be factored out in the integral shown in Eq.~(\ref{eq-pexc_2}) and takes the value corresponding to the equilibrium geometry of the ground state ($R_{eq}\sim1.094$ \AA). As can be seen, the time delay difference between the $v'=1$ and $v'=0$ levels obtained within the Franck-Condon approximation is nearly zero for all photon energies, failing to reproduce the theoretical results accounting for nuclear motion (black circles) and, consequently, the experimental measurements shown in Fig. 3(c).

\begin{figure*}[!htb]
    \centering
    \includegraphics[width=1.0\textwidth]{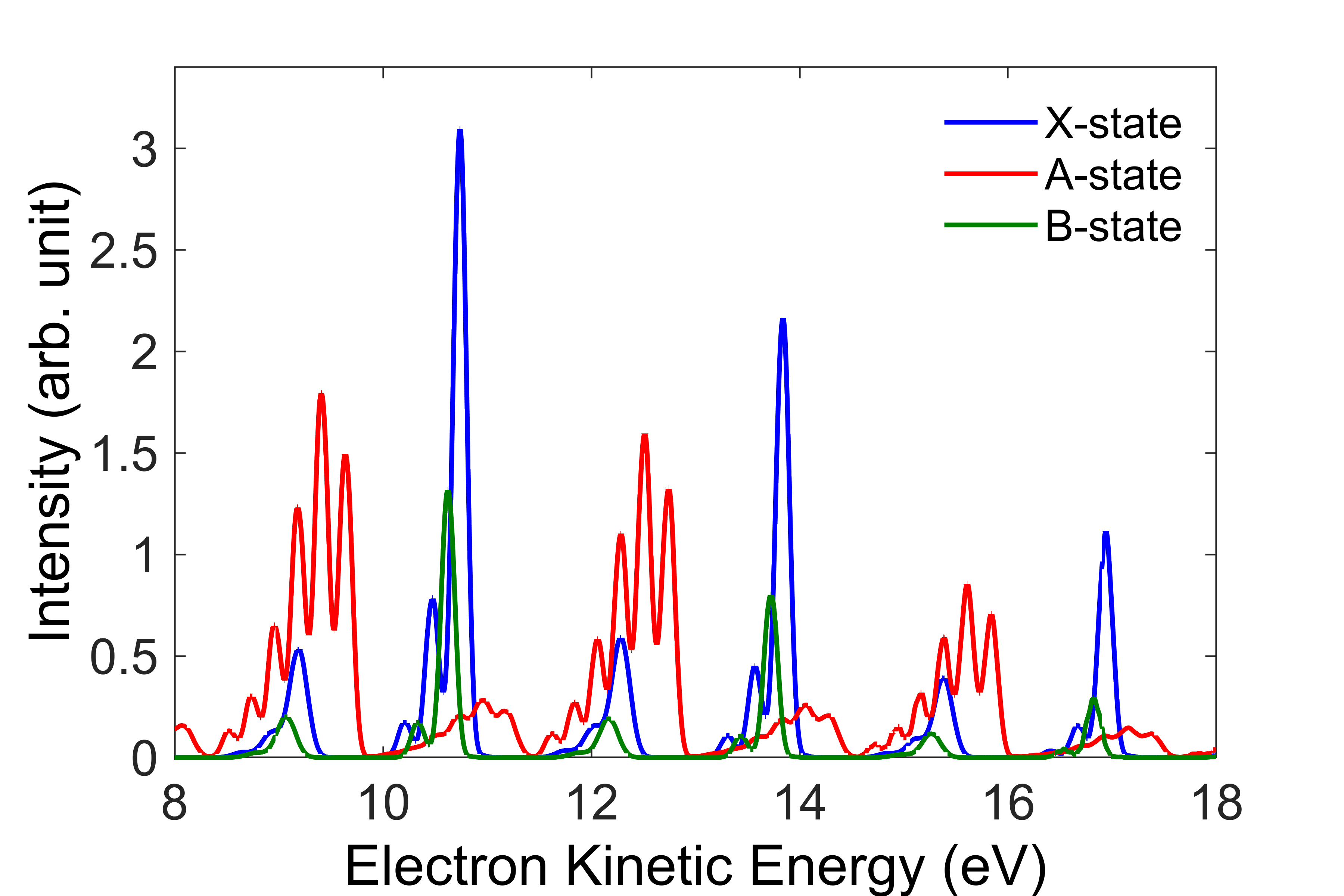}
    \caption*{ \textbf{Fig.S1} Theoretical XUV+NIR photoelectron spectra for the first three excited states: $X$, $A$ and $B$ in N$_2^+$ ion.}
    \end{figure*}

\begin{figure}[htb!]
\begin{center}
\hbox{
\vbox{
\includegraphics[width=0.75\textwidth]{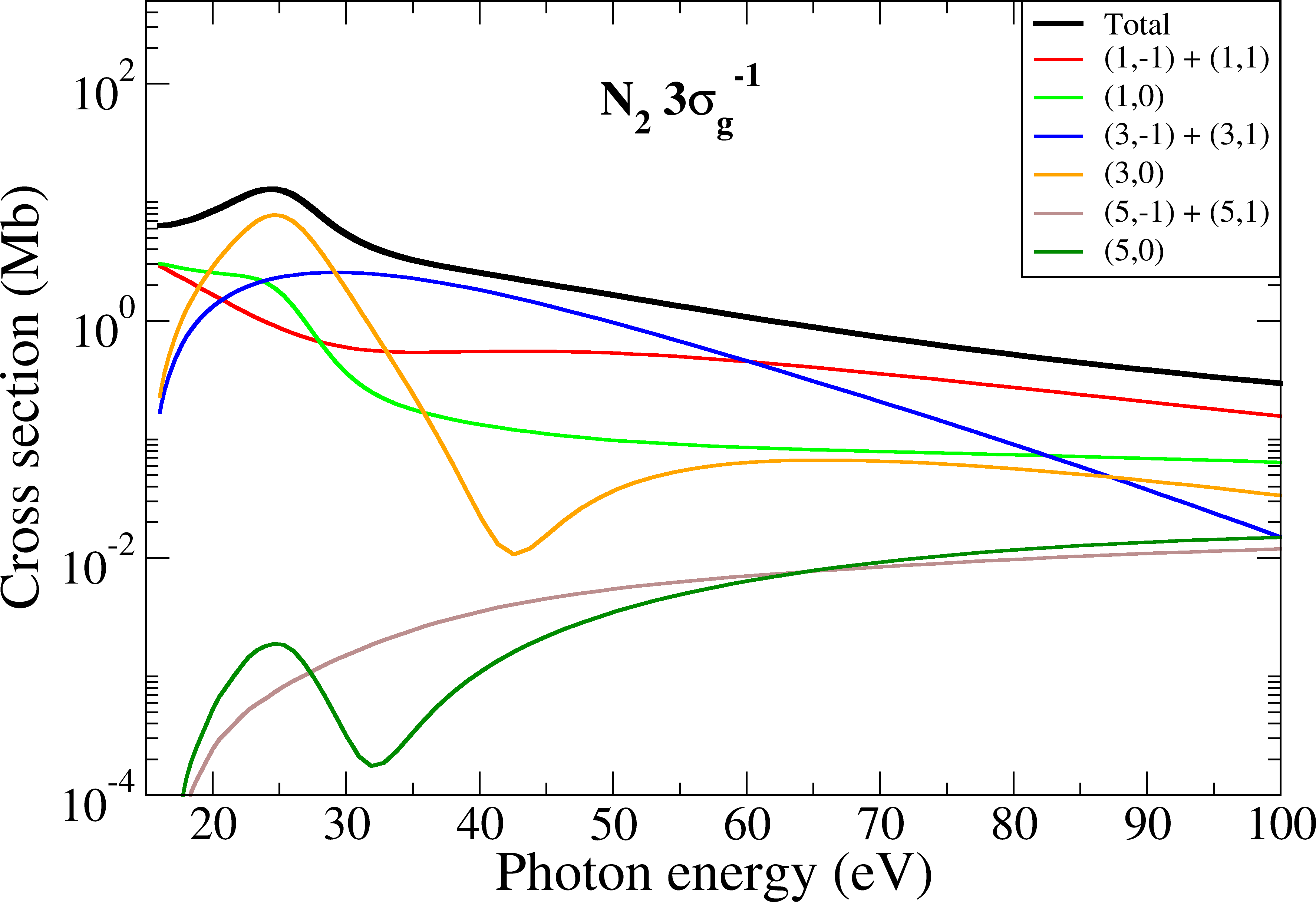}
\vskip 0.25in
\includegraphics[width=0.75\textwidth]{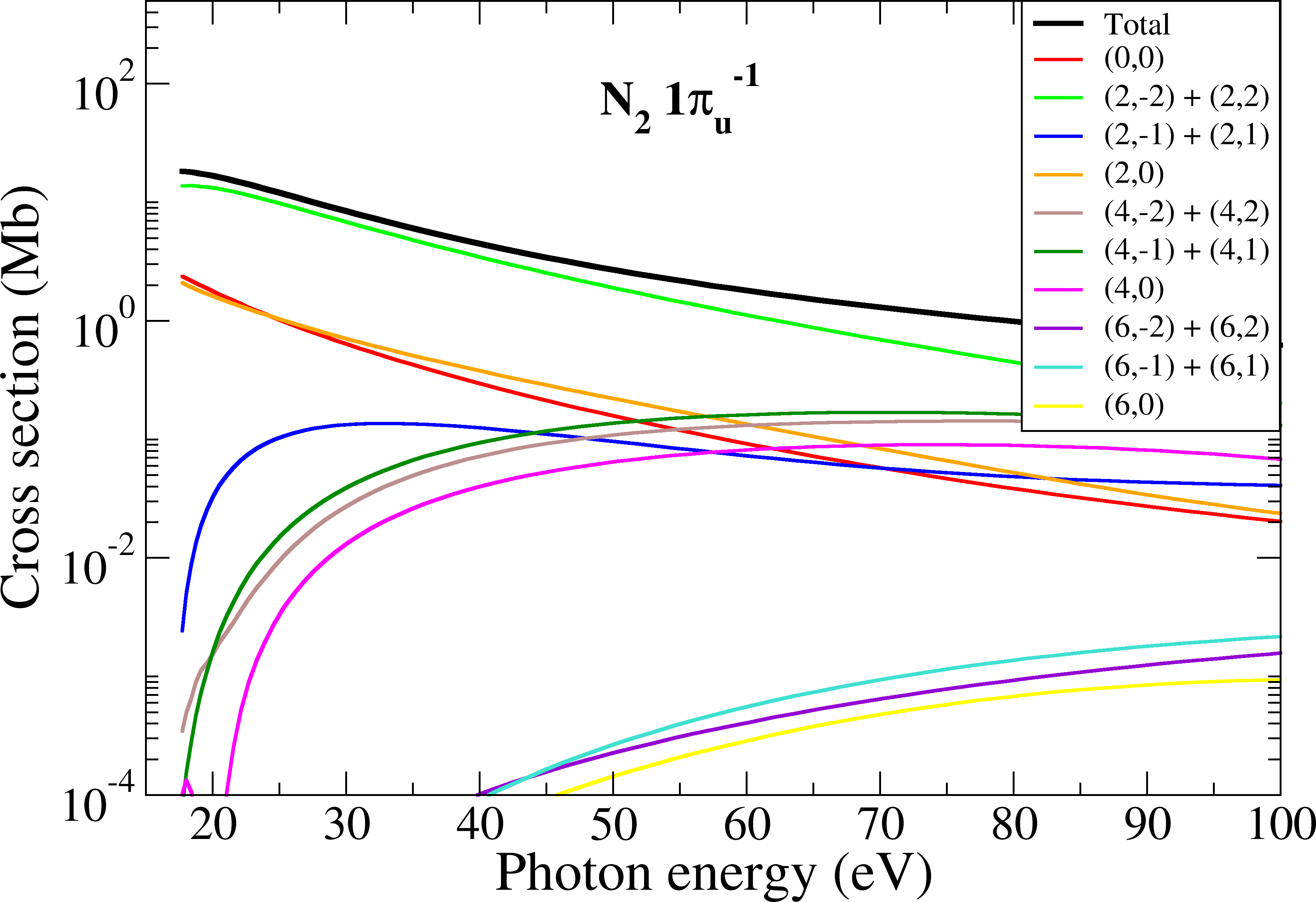}
}
}
\end{center}
\caption*{\textbf{Fig.S2} Differential cross-sections for photoionization of N$_2$, for the $X$ (top) and $A$ (bottom) final states. The angular momentum channels are indicated by ($\ell,m$).}
\end{figure}

\begin{figure*}[!htb]
    \centering
    \includegraphics[width=0.75\textwidth]{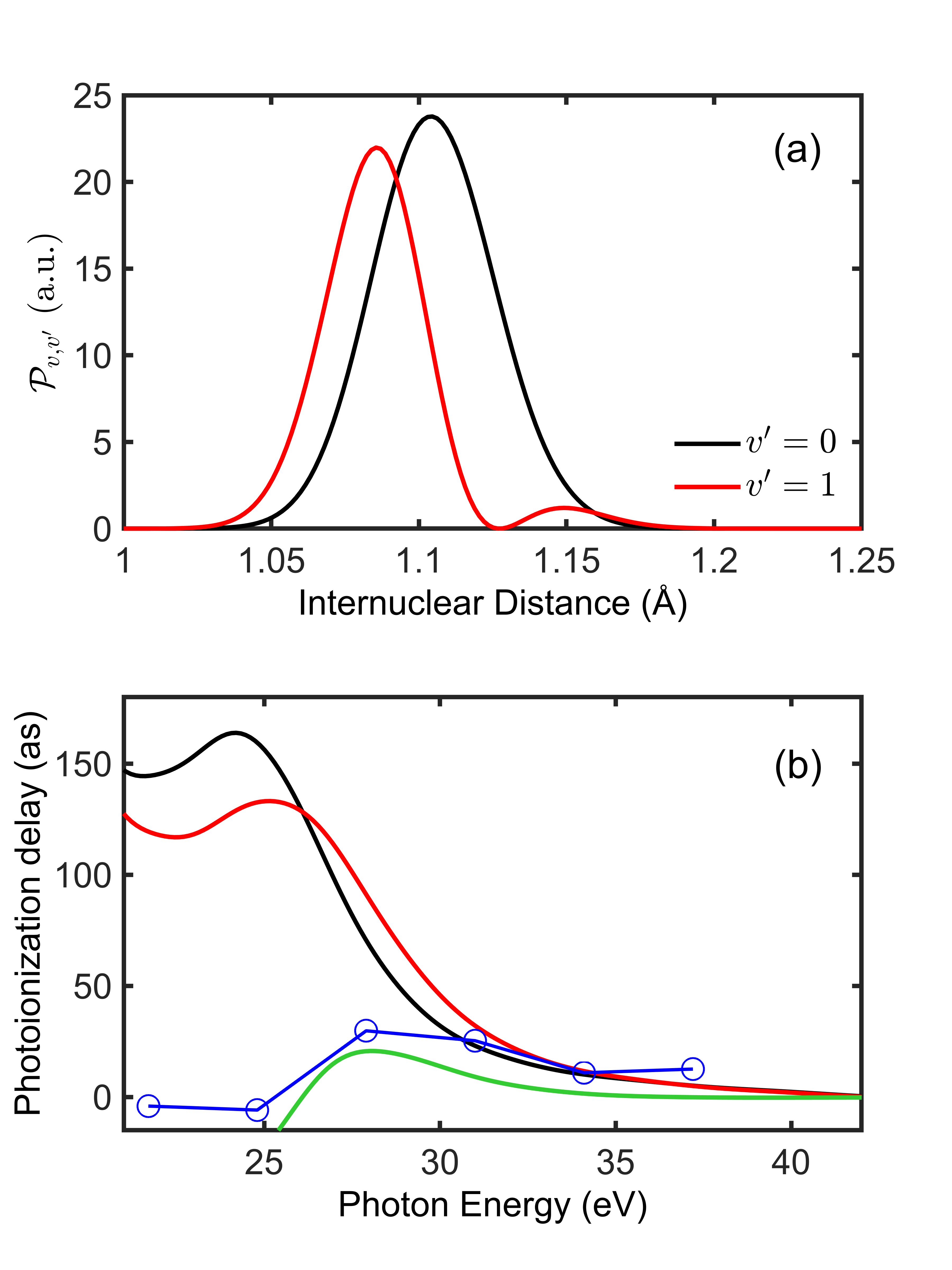}
    \caption*{\textbf{Fig.S3} Same as in Fig.5 extracted for a higher photoelectron energy of $11.7$ eV. (a) Absolute square of the transition matrix element at an electron kinetic energy of $11.7$ eV, multiplied by the initial and final vibrational wave functions, $\mathcal{P}_{v,v'}$, as a function of the internuclear distance $R$. The red (black) curve corresponds to the transition from the $v=0$ level in the ground state of the neutral molecule to the $v'=1$ ($v'=0$) level of the $X$-state in N$_2^+$ ion. (b) The blue circles represent the two-photon relative photoionization delay between $v'=1$ (at $R=1.084$ \AA) and $v'=0$ (at $R=1.103$ \AA) of the $X$-state, extracted from the simulated RABBIT spectrograms at these two $R$-values.}
    \end{figure*}

\begin{figure*}[!htb]
    \centering
    \includegraphics[width=1.0\textwidth]{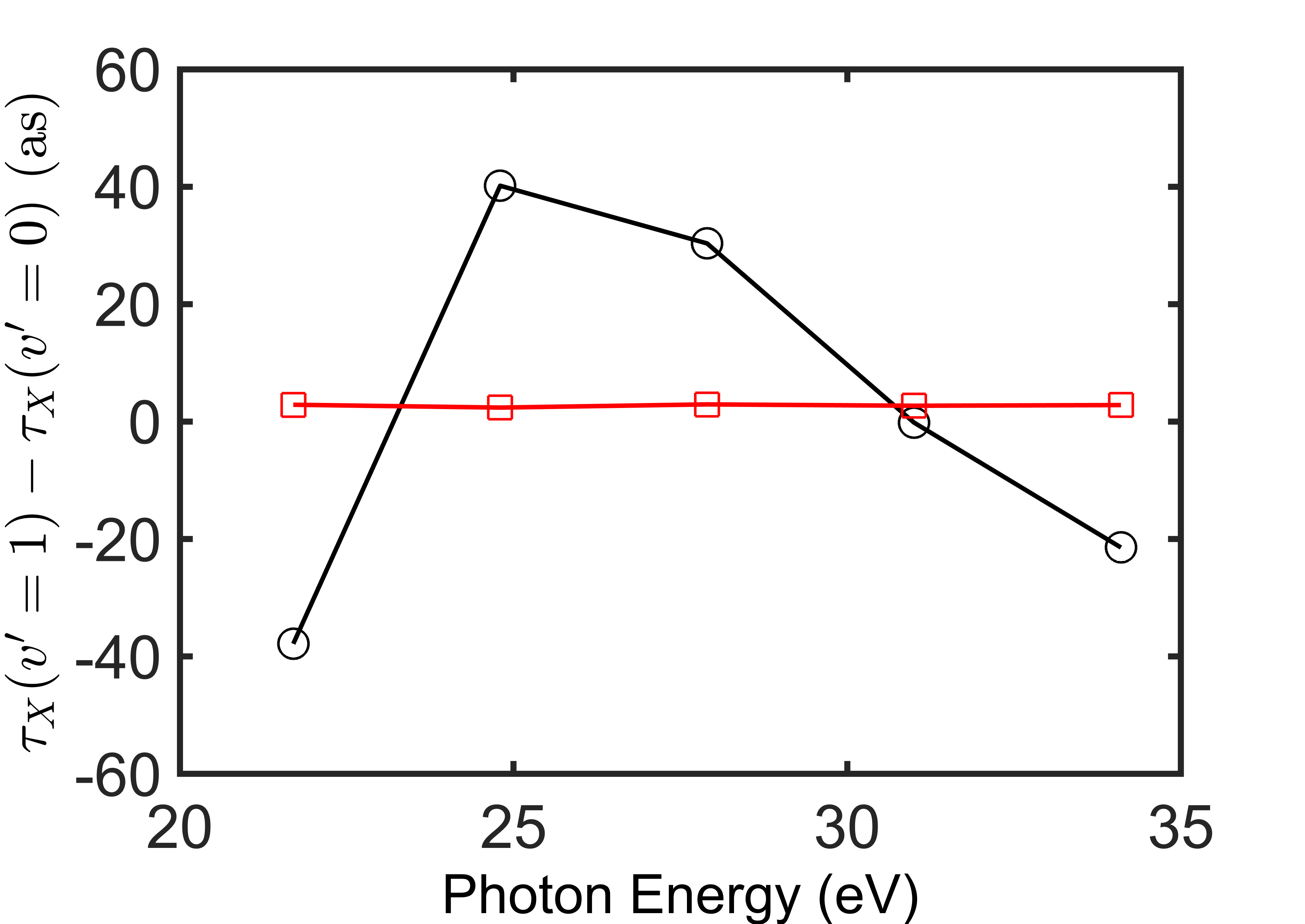}
    \caption*{\textbf{Fig.S4} 
Theoretical relative time delays between the vibrational levels $v'=1$ and $0$ for the $X$-state in N$_2^+$ ion.  Black circles: time delays corresponding to the phases extracted from a RABBIT spectrum in the full calculation ($\tau_{mol}$) as already shown in Fig.3(c) in the main manuscript. Red squares: Results obtained using the Franck-Condon approximation.}
\end{figure*}